\documentclass[sigplan,screen]{acmart}
\AtBeginDocument{%
  \providecommand\BibTeX{{%
    \normalfont B\kern-0.5em{\scshape i\kern-0.25em b}\kern-0.8em\TeX}}}

\setcopyright{acmcopyright}
\acmPrice{15.00}
\acmDOI{10.1145/3471871.3472961}
\acmYear{2021}
\copyrightyear{2021}
\acmSubmissionID{icfpws21erlangmain-p1-p}
\acmISBN{978-1-4503-8612-8/21/08}
\acmConference[Erlang '21]{Proceedings of the 20th ACM SIGPLAN International Workshop on Erlang}{August 26, 2021}{Virtual Event, Republic of Korea}
\acmBooktitle{Proceedings of the 20th ACM SIGPLAN International Workshop on Erlang (Erlang '21), August 26, 2021, Virtual Event, Republic of Korea}

\usepackage{float}
\newfloat{lstfloat}{htbp}{lop}
\floatname{lstfloat}{Listing}

\usepackage{listings}
\usepackage{xcolor}
\definecolor{codegreen}{rgb}{0,0.6,0}
\definecolor{codegray}{rgb}{0.5,0.5,0.5}
\definecolor{codepurple}{rgb}{0.58,0,0.82}
\definecolor{backcolour}{rgb}{0.95,0.95,0.92}
\lstdefinestyle{ourerlangstyle}{
    language=erlang,
    commentstyle=\color{codegreen},
    keywordstyle=\color{magenta},
    numberstyle=\tiny\color{codegray},
    stringstyle=\color{codepurple},
    basicstyle=\ttfamily\footnotesize,
    breakatwhitespace=false,         
    breaklines=true,                 
    captionpos=b,                    
    keepspaces=true,                 
    numbers=left,                    
    numbersep=5pt,
    xleftmargin=10pt,
    framexleftmargin=10pt,
    showspaces=false,                
    showstringspaces=false,
    showtabs=false,                  
    tabsize=2
}
\acmSubmissionID{???-???-???}
\usepackage{microtype}
\begin{document}

\title{Detecting Oxbow Code in Erlang Codebases with the Highest Degree of Certainty}

\author{Fernando Benavides Rodríguez}
\email{elbrujohalcon@gmail.com}
\affiliation{%
  \institution{NextRoll Inc.}
  \city{Sant Cugat del Vallès}
  \country{Spain}}

\author{Laura M.~Castro}
\orcid{0000-0002-3028-1523}
\email{lcastro@udc.es}
\affiliation{%
  \institution{ICT Research Centre -- Universidade da Coruña}
  \streetaddress{Universidade da Coruña}
  \city{A Coruña}
  \country{Spain}
}

\begin{abstract}
The presence of source code that is no longer needed is a handicap to project maintainability. The larger and longer-lived the project, the higher the chances of accumulating dead code in its different forms. 

Manually detecting unused code is time-consuming, tedious, error-prone, and requires a great level of deep knowledge about the codebase. 
In this paper, we examine the kinds of dead code (specifically, oxbow code) that can appear in Erlang projects, and formulate rules to identify them with high accuracy.
  
We also present an open-source static analyzer that implements these rules, allowing for the automatic detection and confident removal of oxbow code in Erlang codebases, actively contributing to increasing their quality and maintainability.

\end{abstract}



\maketitle

\section{Introduction}
Large codebases tend to become even larger and harder to maintain over time. Systems rarely remain stable, and neither do the teams that manage them. As systems grow, what also grows is the complexity of implementing every desired change, reducing at the same time the developers' confidence when applying such changes~\cite{Morgenthaler20121,Scanniello2014392}.
In most teams, in most systems, there comes a time when developers have to make a change and, giving the high uncertainty they have on the code they are working with, they force themselves to do \textit{the smallest change possible}.

Particularly when removing some functionality, developers might be tempted to just remove the particular function or the particular expression that implements the functionality, leaving behind other pieces of code (like extra arguments, macros, environment variables, constants, etc.) that are not used anymore. These elements that remain in the code but are not used in production are what we call \textbf{oxbow code}.

Oxbow code is source code that is not used, generally not tested, sometimes not even compiled, but it stays in the codebase, increasing the complexity and slowly but surely taking a toll on the system's  maintainability.
Detecting the presence of oxbow code, in general, is not a trivial task. In the particular case of Erlang, some examples of oxbow code instances are:

\begin{itemize} \label{oxbow_examples}
\item macros that are no longer used
\item environment variables that are no longer consumed
\item header files that are no longer used or that are used by a single module (i.e.~they do not need to be shared)
\item record fields that are no longer necessary
\item function arguments that can be removed since they are ignored in all the function clauses
\end{itemize}

However, listing all possible oxbow code situations is not enough to solve the problem. Consider an application with the following configuration:

\begin{lstlisting}
{application,
 lapp,
 [{description, "A large application"},
  {vsn, "10.3.142"},
  {env, [{sample_rate, 0.5}]}]}.
\end{lstlisting}

And consider that we also have a module \verb`lapp` like:

\begin{lstlisting}
-module(lapp).
-include_lib("eunit/include/eunit.hrl").
-define(DEFAULT_SAMPLE_RATE, 0.25).
-export([run/1]).

run(Param) ->
  SampleRate =
    application:get_env(
      lapp, sample_rate,
      ?DEFAULT_SAMPLE_RATE),
  Result =
    maybe_evaluate(
      SampleRate, Param),
  logger:info(#{param => Param,
                result => Result}),
  Result.

maybe_evaluate(SampleRate, Param) ->
  maybe_evaluate(
    rand:uniform(), SampleRate, Param).

maybe_evaluate(Rand, Rate, _Param)
  when Rand >= Rate ->
    ignored;
maybe_evaluate(_Rand, _Rate, Param) ->
  evaluate(Param).

evaluate(Param) ->
  {evaluated, Param}.

%% unit tests
ignore_test() ->
  application:set_env(
    lapp, sample_rate, 0.0),
  ?assertEqual(ignored, run(anything)).

evaluate_test() ->
  application:set_env(
    lapp, sample_rate, 1.0),
  ?assertEqual(
    {evaluated, anything},
    run(anything)).
\end{lstlisting}

When someone uses the exported function \verb`lapp:run/1`, which in turn invokes private function \verb`maybe_evaluate/2`, this implementation will sometimes run \verb`evaluate/1` (see line 26) and sometimes it will not (see line 24), based on a sampling limited by the configuration parameter \verb`sample_rate` (see line 20). These behaviors are exercised by test functions \verb`ignore_test/0` and \verb`evaluate_test/0` respectively.

If a given developer is now tasked with removing that sampling, that is to say, if the call to \verb`evaluate/1` became \textit{mandatory} and the need for sampling eliminated, how would they proceed?

To modify the smallest amount of code possible, and following a test-driven development approach (TDD), the first step would be to remove \verb`ignore_test/0` (lines 32--35) and the setting of the \verb`sample_rate` variable in \verb`evaluate_test/0` (lines 38--39). That will result in \verb`evaluate_test/0` failing and, to fix it, the developer would have to remove the first clause of \verb`maybe_evaluate/3` (lines 22--24).
That is enough to fix the failing test and complete the task, but it leaves many lines of code that serve no purpose at all anymore.

First, \verb`maybe_evaluate/3` does no longer need to receive three arguments, since it is ignoring the first two. As a matter of fact, the function itself is not needed either, for its only purpose is now to call \verb`evaluate/1` with the same parameter it receives. The developer can just call \verb`evaluate/1` in \verb`maybe_evaluate/2`:

\begin{lstlisting}[firstnumber=21]
maybe_evaluate(_SampleRate, Param) ->
  evaluate(Param).
\end{lstlisting}

This leads to the same situation repeating: function \texttt{maybe\_\-evaluate/2} is also superfluous now. The developer can call \verb`evaluate/1` directly from \verb`run/1`:

\begin{lstlisting}[firstnumber=9]
run(Param) ->
  SampleRate =
    application:get_env(
      lapp, sample_rate,
      ?DEFAULT_SAMPLE_RATE),
  Result = evaluate(Param),
  logger:info(#{param => Param,
                result => Result}),
  Result.
\end{lstlisting}

At this stage, \verb`SampleRate` is no longer needed and the compiler emits a warning about that, which the developer gets rid of by rewriting \verb`run/1` once more:

\begin{lstlisting}[firstnumber=9]
run(Param) ->
  Result = evaluate(Param),
  logger:info(#{param => Param,
                result => Result}),
  Result.
\end{lstlisting}

Finally, the developer can safely remove both the macro \verb`?DEFAULT_SAMPLE_RATE` from \verb`lapp.erl` and the environment parameter \verb`sample_rate` from \verb`lapp.app.src`:

\begin{lstlisting}
{application,
 lapp,
 [{description, "A large application"},
  {vsn, "10.3.142"}}.
\end{lstlisting}

\begin{lstlisting}
-module(lapp).
-include_lib("eunit/include/eunit.hrl").
-export([run/1]).

run(Param) ->
  Result = evaluate(Param),
  logger:info(#{param => Param,
                result => Result}),
  Result.

evaluate(Param) ->
  {evaluated, Param}.

evaluate_test() ->
  ?assertEqual(
    {evaluated, anything},
    run(anything)).
\end{lstlisting}

The code is now far smaller than the intermediate version, it performs less unnecessary operations (for instance, the random number generation), and it does not contain anything that is no longer needed in production.

In smaller pieces of code like this example, experienced developers aided by good tests can probably detect all the pieces that should be removed by themselves. Advanced IDEs and Language Servers (like \verb`erlang-ls`~\cite{erlangls}) can also help. However, it is not a trivial task and it becomes far less trivial when systems, modules, and functions are larger and more complex, likely spreading a single feature among multiple files in different folders.

Thorough code reviews can help detect oxbow code, but that will only work if the reviewers have a deep knowledge of the system that they are reviewing. That is because, most of the time, to determine that some piece of code is actually dead, the reviewers must know for sure that it is not used in the rest of the code, \textit{outside} the scope of the changes that they are reviewing. In fact, given the complexity of this process, it is generally safer to \textit{just let the oxbow code be untouched}, although of course not practical and sustainable in the long run. This is what brings up the need for an automatic tool to detect oxbow code.

Such tool should be able to analyze all the modules, headers, and other files in an Erlang system and, based on some rules that characterize the oxbow code instances to look for (like the ones listed on page~\ref{oxbow_examples}), detect code elements that are not needed anymore and therefore can be (and should be) confidently removed. For the sample code we just used, we could expect to see the following diagnosis\footnote{Although presented here all together for simplicity, this output corresponds to the different interaction steps that would mimic the manual ones we have just described.}:

\begin{lstlisting}[language=erlang,basicstyle=\tiny]
% After removing the first clause of maybe_evaluate/3
src/lapp.erl:18: maybe_evaluate/3 does not need its #2 argument
src/lapp.erl:18: maybe_evaluate/3 does not need its #1 argument

% After removing maybe_evaluate/3 altogether
src/lapp.erl:15: maybe_evaluate/2 does not need its #1 argument

% After removing maybe_evaluate/2 altogether
src/lapp.erl:5: ?DEFAULT_SAMPLE_RATE is unused
src/lapp.app.src:0: sample_rate is not used anywhere in the code
\end{lstlisting}

In this paper, we present a taxonomy of types of unused code in Sect.~\ref{sec:taxonomy}, and contribute with a set of precise rules that can be used to detect well-defined examples of each of them in Erlang. Then, in Sect.~\ref{sec:automation}, we analyze how those rules can be automated as part of a code analysis tool both in theory and in practice (i.e.~with the existing syntactic tools provided by Erlang/OTP). Finally, we present \textbf{Hank}~\cite{hank}, an automatic oxbow code detector that currently supports a subset of those rules, in Sect.~\ref{sec:hank}, and shows concrete examples that demonstrate its features and capabilities.

\section{Characterisation of Unused Code} \label{sec:taxonomy}
The most common and informal term to refer to \textit{unused} source code is \textit{dead code}. However, there are several nuances that we might imply or not when referring to this concept. These are captured by the more specific terms:

\begin{description}
\item [Unreachable Code] Part of the source code of a program which can never be executed because there exists no control flow path to the code from the rest of the program. In Erlang, these situations may be detected using tools like XRef~\cite{xref} and Dialyzer~\cite{dialyzer}.
\item [Dead Store] A local variable that is assigned a value but is not read by any subsequent instruction. The Erlang compiler usually emits a warning about these variables.
\item [Redundant Code] Source code or compiled code in a computer program that is unnecessary, such as recomputing a value that has previously been calculated and is still available.
\item [Oxbow Code] Fragments of program code that were once needed but which are now never used.
\end{description}

The main focus of this work is this last category, i.e.~\textbf{oxbow code}. Oxbow code is typically the result of program modification, either when an item is superseded with a newer version but the old version is not removed, or when an item is removed or replaced, but the item's supporting code is not.

While oxbow code cannot be the source of bugs in itself (for it is never executed), it reduces the system's maintainability by making it harder for developers to apply changes. Especially for junior developers or simply developers unfamiliar with the code, they typically cannot be sure that the dead code is, in fact, unused.

\subsection{Oxbow Code in Erlang Projects}
There are many variants of oxbow code that can be seen in Erlang source files (as those listed in the previous section), and before our work, there were no tools to spot them accurately.
In this subsection, we will present the analysis we did on several relevant Erlang projects and the rules for detecting them that we designed on the basis of the examples of oxbow code that were present on them.

\subsubsection{Existing Code Analysis}
Mostly due to how oxbow code appears, it is far more prevalent in large and old systems. We analyzed several large open-source Erlang codebases, namely:

\begin{description}
    \item [Erlang/OTP~\cite{erlang}] As the oldest Erlang codebase, the own implementation of the language and its standard libraries is the most interesting and relevant project to consider.
    \item [Kazoo~\cite{kazoo}] A software platform designed to provide car\-rier-grade VoIP features. As one of the largest open-source Erlang repositories in existence, it is a relevant project to check. Furthermore, it being an umbrella project means it is self-contained.
    \item [MongooseIM~\cite{mongoose}] Another very large Erlang system, in this case, an instant messaging platform aimed at large installations. The meticulous maintenance that it is subjected to makes it another excellent candidate to analyze. We do not expect to find a large number of oxbow code instances here.
\end{description}

\begin{table*}
  \caption{Oxbow code presence in several Erlang projects}
  \label{tab:analysis}
  \begin{tabular}{lccc}
  \toprule
   & \textbf{Erlang/OTP} & \textbf{Kazoo} & \textbf{MongooseIM} \\
  \midrule
  Size of project (in SLOC) & $> 2M$ & $> 400K$ & $> 180K$ \\
  Number of active maintainers & $10$ & $6$ & $5$ \\
  Project age (in years) & $> 20$ & $11$ & $8$ \\
  \midrule
  Unused macros                 & $> 1000$ & $> 150$ & $90$ \\
  Unused record fields          & $> 200$  & $> 20$  & $7$ \\
  Unused header files           & $> 15$   & $3$     & \\
  Unused configuration options  & $1$      & $3$     & \\
  Unused function arguments     & $> 1000$ & $> 500$ & $> 200$ \\
  \bottomrule
  \end{tabular}
\end{table*}

Table~\ref{tab:analysis} summarizes our findings in terms of oxbow code presence in these projects. With numbers like these in mind, and even if just a fraction of them were actually caused by oxbow code, the need for a tool that can confidently detect these elements with high accuracy becomes evident.

\subsubsection{Oxbow Code Classification}
Oxbow code, particularly in Erlang, can present itself in multiple ways. Therefore, one of our first tasks was to identify as many of them as possible. Using the projects listed above and our personal experience, we generated the following non-exhaustive list of oxbow code classes:
\begin{description} \label{oxbow_classes}
    \item [Unused Macros] These are macros that are defined using a \verb`define` attribute but not used anywhere in the code. They are usually the result of the removal of some code where they were originally used.
    \item [Unused Record Fields] These are record fields that are declared using a \verb`record` attribute and then never set nor accessed anywhere in the code. This situation may arise when the last place where a particular record field is used is removed, but the corresponding field definition is left in place.
    \item [Unused Header Files] These are files (usually with \verb`hrl` extension within an \verb`include` folder) that are supposed to be used in one or more Erlang modules, but that is not the case. The existence of such files is usually the result of removing the modules where they were included or just the \verb`include` attributes in them, without removing the associated files.
    \item [Unused Configuration Options] These are application configuration parameters (i.e.~the ones that are retrieved using \verb`application:get_env/2,3`) that have a value but are never consumed throughout the code. Like the previous examples, they may have been consumed once, but the code that consumed them was removed, and they remained in \verb`app.src` or \verb`sys.config` files.
    \item [Unused Function Arguments] These are positional arguments that are consistently ignored in all the clauses of a function. Except when the functions are called dynamically and are expected to have a fixed arity (e.g.~if they are callbacks for a particular behavior), these functions can serve their purpose without such arguments.
\end{description}

\section{Tool Design} \label{sec:automation}
Our goal was to build a tool that could detect all the classes of oxbow code listed in the previous section, as well as new ones that can be defined in the future, automatically and accurately. These were its desired features:
\begin{itemize}
    \item Ease of \textbf{use}.
    \item Detailed \textbf{configuration}.
    \item Dialyzer-style levels of \textbf{certainty}.
    \item With room for \textbf{extensibility}.
\end{itemize}

\subsection{Ease of Use}
For developers to be able to, and more importantly, \textit{willing} to use a brand-new tool, this tool should not require complex configuration or set up~\cite{Nilsson2015}. At least, not to get the basic stuff running. It is also essential that the results the tool produces are clear and actionable.

In this regard, the Erlang ecosystem already has an extensible plugin \textit{framework} for tools to run: \verb`rebar3`~\cite{rebar3}. There are also other existing tools with similar interfaces (like \verb`rebar3_lint`~\cite{rebar3lint} or \verb`rebar3_format`~\cite{rebar3format}) that we can use for inspiration and be consistent with behavior that is already intuitive to the Erlang developers around the world.

It is also important to consider that a tool like \textbf{Hank} is helpful to be run on the developer's computer, but it is far more useful if it is set up to run as a \textit{continuous integration} task that validates code automatically. That is why it was very important for this tool to be easy to integrate into any CI pipeline and \verb`rebar3` helps enormously with that.

Effectively, running \textbf{Hank} on a rebar3-maintained project is as simple as adding an entry to the \texttt{project\_\-plugins} list in \verb`rebar.config` and executing \verb`rebar3 hank` in the shell.
That will analyze all source files in the project with all the default rules that \textbf{Hank} provides to find different kinds of oxbow code. Then it will produce a list of warnings with filenames and line numbers that the developers can follow to investigate and eventually remove or refactor.

\subsection{Detailed Configuration}
For tools to be run as checks in CI pipelines, they need to be \textit{certain} (i.e.~they need to produce an exit code other than \verb`0` if and only if they find errors) and thus they should allow developers to specify the errors they want to ignore or the files they do not want to analyze.

That is why we made \textbf{Hank} extensively configurable. Its default configuration should analyze all files and report all warnings, but it provides several different ways to ignore particular warnings, particular files, or even entire checks.

\subsection{Full Certainty}
One of the foundational goals of \textbf{Hank} is to allow developers to remove code without fear. That can only be achieved if the tool produces no \textbf{false positives}. Otherwise, the developers will always doubt if what the tool reports as a removable piece of code can \textit{actually} be removed, and they will keep it \textit{just in case}. We would be back to square one.

Nevertheless, as we will explain in the next section, most rules cannot be implemented with full accuracy. Even for rules that are fully defined in theory, implementing an algorithm that would detect all instances of that kind of oxbow code and nothing else can be extremely complex, hard on performance, etc. There is a trade-off to make:

\begin{itemize}
    \item the algorithm may report all the instances of certain kind of dead code \textit{and some other things that are not actual instances of it} (i.e.~false positives) or,
    \item the algorithm may report only instances of that kind of oxbow code, but maybe \textit{not all of them}.
\end{itemize}

The latter is how Dialyzer~\cite{Sagonas200737} works: it may not report \textit{all} discrepancies in your code, but it is \textbf{never wrong}. In other words, if Dialyzer produces a warning, then developers can be completely sure that there is a discrepancy in their code.

Considering our goal above, we decided that \textbf{Hank} should be built with a \textit{Dialyzer-style} of certainty and avoid false positives even if that means failing to detect some instances of oxbow code. This is indeed the only way of guaranteeing that it is always safe to remove the code pointed out by \textbf{Hank}. 

\subsection{Extensibility}
The list of oxbow code classes from Sect.~\ref{oxbow_classes} is by no means complete. Moreover, certain systems can have their own kinds of oxbow code, and we wanted \textbf{Hank} to be able to detect those as well.

That is why we designed \textbf{Hank} to work with a set of \textit{rules} (one for each class of oxbow code) and made that behavior available to the users, so they can define and run their own rules to detect their particular versions of oxbow code, too.

\section{Hank} \label{sec:hank}
In this section, we will discuss \textbf{Hank} in detail, from its architectural details to the current set of rules and the mechanism that allows developers to define their own. In following sections we will discuss the caveats of the implementation of each rule and how we dealt with those.

\subsection{General Architecture} \label{sec:arch}
As we have already mentioned, \textbf{Hank} is a \verb`rebar3` plugin, and as such it is built as an OTP application, but without a main supervisor. It provides an entry point for the \verb`rebar3` framework in the main module of the app: \verb`rebar3_hank`. As usual with this kind of plugins, the only responsibility of that module is to implement \verb`init/1`, a function that receives the current state from \verb`rebar3` and calls an internal implementation (\verb`rebar3_hank_prv:init/1`) with it.

\verb`rebar3_hank_prv` defines the interface between \textbf{Hank} and \verb`rebar3` and abstracts the rest of the system from its details. The actual implementation of our tool begins in the module called \verb`hank`, which is the facade of the system and is called from \verb`rebar3_hank_prv` after parsing all the input received from \verb`rebar3`.

\begin{figure*}
  \centering
  \includegraphics[width=0.9\textwidth]{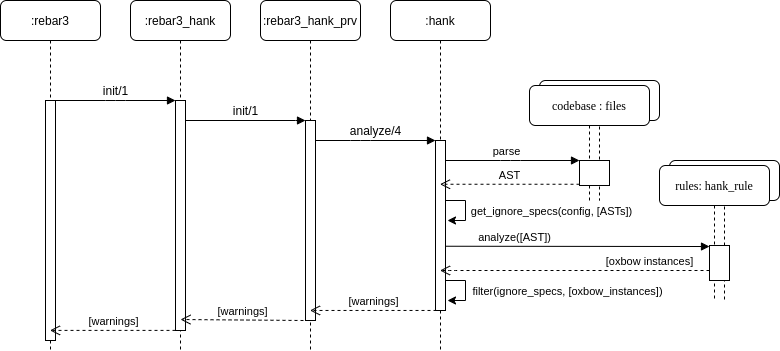}
  \caption{General behavior of \textbf{Hank}}\label{fig:sequence}
  \Description{A UML sequence diagram showing the general behavior of \textbf{Hank}}
\end{figure*}

The \verb`hank` module provides a single point of entry via the \verb`analyze/4` function. This function will traverse all the code in the system under analysis, apply all the rules and return a possibly empty list of warnings with all the oxbow code instances found. An illustration of this process we have just described can be visualized in Fig.~\ref{fig:sequence}.
Specifically, the implementation of \verb`analyze/4` can be summarized as follows:

\begin{enumerate}
    \item Parse all project files and collect their abstract syntax trees (ASTs), which will later be used by the rules to detect instances of oxbow code.
    \item Generate a list with all the ignore specifications. This is done at this point since the list of elements to ignore can be defined in \verb`rebar.config`, but also using the \verb`hank` attribute in the files themselves (i.e.~it will be found in their ASTs).
    \item Run each of the rules against all the files and collect all the oxbow code instances they detect.
    \item Remove the instances that match the ignore specifications, format the remaining ones as warnings, and output that list.
\end{enumerate}

As can be seen in this general algorithm, each rule is run independently from the others. Each one is defined in its module, implementing a shared behavior called \verb`hank_rule`, that defines two callbacks:

\begin{itemize}
    \item \verb`analyze/2` that receives a list of ASTs and returns a list of oxbow code instances.
    \item \verb`ignored/2` that allows each rule to define its way to specify ignored oxbow code instances within modules.
\end{itemize}

This is how \textbf{Hank} provides extensibility: to define a new rule, the only thing a developer needs to do is implement this behavior.

\subsection{Default Rules}\label{defaultrules}
By far, the most interesting part of the system, and the one that required the most effort, is the implementation of each of the rules defined in Sect.~\ref{oxbow_classes}. In this section, we will formalize each of those rules, describing what an \textit{ideal implementation} of such a rule would involve. However, when such an implementation is not feasible, we will exhibit the problems that we found, analyze possible solutions, and describe the actual implementation.

\subsubsection{Unused Macros}\label{unusedmacros}
This rule aims at detecting all instances of macro definitions that are not used anywhere in the code. More specifically, the goal is to find all \verb`define` attributes with no corresponding \verb`macro` nodes anywhere.

As an example, in the following module, the rule should report \verb`UNUSED` as an unused macro, but it should not emit a warning for \verb`USED` since it is used in \verb`my_function/0`:

\begin{lstlisting}
-module(my_module).
-export([my_function/0]).

-define(UNUSED, unused).
-define(USED, used).

my_function() -> ?USED.
\end{lstlisting}

In AST terms, the algorithm behind this rule can be summarized as:
\begin{enumerate}
    \item Find all nodes with \verb`type == attribute` and \texttt{name == define}.
    \item For each one of them, extract the corresponding macro name and arity. We need the arity information because in Erlang, it is valid to have two macros with the same name and different number of arguments, same as happens with function definitions.
    \item Find all nodes with \verb`type == macro`.
    \item For each of them, extract the corresponding macro name and arity.
    \item Return an oxbow code instance for each macro in the first list but not in the second one.
\end{enumerate}

\subsubsection{Unused Record Fields}\label{unusedrecords}
The idea behind this rule is to detect record fields that are defined but not used anywhere in the code. \textbf{Hank} will be looking for \verb`record` attributes that include fields that appear nowhere else in the code.

Consider the following example:

\begin{lstlisting}
-module(my_module).
-export([new/0, use/1]).

-record(my_record, {used, unused}).

new() -> #my_record{}.

use(#my_record{used = Value}) -> Value.
\end{lstlisting}

Upon analyzing this module, the rule should report \verb`unused` as an unused field, but not \verb`used` since it is used in \verb`use/1`.

According to OTP's \verb`erl_syntax` module, there are three types of AST nodes that represent usage of field records:

\begin{description}
    \item[record\_access] A record field access expression, for instance \verb`Var#record.field`.
    \item[record\_expr] A record expression. It can be either like \verb`#record{field = Value, …}` or it can include a variable, like \verb`Var#record{field = Value, …}`.
    \item[record\_index\_expr] A record field index expression, for instance \verb`#record.field`.
\end{description}

Considering the list above, in AST terms, the algorithm for this rule can be summarized as:
\begin{enumerate}
    \item Find all nodes with \verb`type == attribute` and \texttt{name == record}.
    \item With them, generate a list of all fully-qualified record field names, like \verb`{RecordName, FieldName}`.
    \item Find all nodes with any of the three types listed above.
    \item With them, build a list of all pairs of fully-qualified record field names.
    \item Return an oxbow code instance for each tuple that is on the first list but not on the second one.
\end{enumerate}

\subsubsection{Unused Header Files}\label{unusedheaders}
This rule's goal is to detect \verb`*.hrl` files that are not included in any module.
To achieve that goal, this rule's algorithm is simple:

\begin{enumerate}
    \item Find all the paths to files with \verb`hrl` extension.
    \item Find all the nodes with \verb`type == attribute` and \texttt{name == include} or \texttt{name == include\_lib}.
    \item Return an oxbow code instance for each file in the first list that does not correspond to any of the paths in the second list.
\end{enumerate}

\subsubsection{Unused Configuration Parameters}\label{unusedconfigs}
For this rule, the idea is to find all application configuration parameters (the ones that can be retrieved via \verb`application:get_env`) that are defined either in \verb`.app.src` or \verb`sys.config` files but never used within the system.

When set up to write down an algorithm for this rule, we immediately find out that finding all places where a certain parameter is used \textit{precisely} (i.e.~without missing any) is not as simple as it seems. It is not enough to find all usages of \verb`application:get_env/1,2,3` with the parameter in the corresponding position. That is because somebody can wrap up that call in a function, like in the following example:

\begin{lstlisting}
-module(my_module).
-export([get_env/1]).

get_env(Param) ->
    application:get_env(my_app, Param, unknown).
\end{lstlisting}

In order to preserve our desired \textit{Dialyzer-style} level of certainty in our results, we decided to count \textit{any} appearance of the configuration parameter as a usage. That will generate some false negatives (undetected instances of oxbow code).

For instance, when \textbf{Hank} analyzes the code above, if there is a parameter called \verb`unknown` in the application configuration that is not used anywhere, the rule will not report it as unused since the atom \verb`unknown` appears somewhere in the code, even if no one ever calls \verb`get_env/1`.

With that in mind, the algorithm we implemented is:

\begin{enumerate}
    \item From \verb`rebar3` context, gather the names of all the applications being built (for a single repository can hold more than one OTP application).
    \item Find all files containing configuration parameters (that is, files with extensions like \verb`config` or \verb`app.src`).
    \item From those files, create a list with all the configuration options associated to the applications listed before.
    \item Create a list with all the nodes with \verb`type == atom` in all the modules and header files being analyzed.
    \item Return an oxbow code instance for each element in the first list that is not a member of the second one.
\end{enumerate}

\subsubsection{Unnecessary Function Arguments}\label{unusedargs}
The idea behind this rule is to detect all instances of positional arguments that are consistently ignored in all the clauses of a function, and therefore can be unnecessary to that function's behavior. An example of such a situation follows:

\begin{lstlisting}
-module(my_module).
-export([external/1]).

external(Param) ->
    internal(Param, some:computation()).

internal(Used, _Unused) when is_integer(Used) ->
    {is_integer, Used};
internal(_Used, _Unused) ->
    not_integer.
\end{lstlisting}

For the module above, the rule should identify \verb`internal`'s second argument as an unnecessary function argument, but it should not report the first argument since it is indeed used in its first clause.
Finding this kind of oxbow code is particularly important since, as in the example, it may effectively lead to the removal of unneeded computations, which may improve the system's performance of the system as well as making it more maintainable.

It is relevant to mention that the algorithm for this rule does not need to look for unused variables. That is to say, it does not need to figure out if \verb`_Unused` is actually unused in the function clause. That is because if it had a non-underscored name (like \verb`Unused`), the compiler itself would emit a warning and, for variables with underscore prefixes, there are other tools (like Elvis/\verb`rebar3_lint`) that will generate a warning if they are, rather incorrectly, used. This means that, to determine that an argument is unused, this rule implementation only needs to check if it is either \verb`_` or an underscored variable (like \verb`_Unused`).

Under this consideration, the resulting algorithm is:

\begin{enumerate}
    \item Find all nodes with \verb`type == function`.
    \item For each of them, find all their clauses.
    \item For each clause, find all the patterns and keep the ones that are just a node with either \verb`type == underscore`, or \verb`type == variable` and a name that starts with \verb`_`.
    \item For each function, return an oxbow code instance for each element that is a member of the list of ignored arguments in all its clauses.
\end{enumerate}

\subsection{Implementation of New Rules}
As we mentioned in Sect.~\ref{sec:automation}, we do not want to limit the number of rules that \textbf{Hank} can apply. On the contrary, \textbf{Hank} is designed to allow developers to build their own rules by implementing the \verb`hank_rule` behavior.
As stated in Sect.~\ref{sec:arch}, the behavior only defines two callbacks: \verb`analyze/2` and \verb`ignored/2`. To understand how to implement a new rule, developers can check any of the existing ones as a source of inspiration or copy the template on Lst.~\ref{src:ruletemplate}.

\begin{lstfloat}
\begin{lstlisting}[caption={Hank rule template.},label={src:ruletemplate}]
%% @doc |A brief description of the rule|.
%%      <p>To avoid this warning, |...|.</p>
-module(my_new_rule).
-behavior(hank_rule).
-export([analyze/2, ignored/2]).

%% @doc |Explain the algorithm and its caveats|
-spec analyze(hank_rule:asts(),
              hank_context:t()) ->
                 [hank_rule:result()].
analyze(FilesAndASTs, Context) ->
    % Traverse the ASTs and collect the nodes
    % that are relevant for the rule analysis
    RelevantNodes =
        [{File, Node}
         || {File, AST} <- FilesAndASTs,
            is_relevant_file(File, Context),
            Node <- relevant_nodes(AST, Context)],

    % Select the ones that are actual oxbow
    % code instances and generate a result
    % for each one of them
    [#{file => File,
       line => 0,
       text => Text,
       pattern => Key} % this matches ignored/2
     || {File, Key, Text}
            <- run_analysis(RelevantNodes)].

%% @doc Should the rule analyze this file?
is_relevant_file(_File, _Context) -> false.

%% @doc Pick the nodes that are actually relevant.
%%      You will want to use erl_syntax,
%%      erl_syntax_lib, and hank_utils to traverse
%%      the AST and build the output list.
relevant_nodes(_AST, _Context) -> [].

%% @doc Actually do the analysis to find
%%      oxbow code instances.
run_analysis(_Nodes) -> [].

%% @doc According to the pattern,
%%      should we ignore this result?
%%      This should match analyze/2 output.
-spec ignored(hank_rule:ignore_pattern(),
              term()) -> boolean().
ignored(_Pattern, _IgnoreSpec) -> false.
\end{lstlisting}
\end{lstfloat}

\section{Discussion}

This section provides further insights into some of the design or implementation decisions that affect \textbf{Hank}.

\subsection{Rule Implementation Caveats}
When trying to implement most, if not all, of the rules in Sect.~\ref{defaultrules}, we found some corner cases where our rules either failed to detect actual oxbow code instances (\textit{false negatives}) or reported as oxbow code instances some pieces of code that were actually used (\textit{false positives}). Our intention was always to remove \textit{false negatives} and \textit{false positives} completely, but in some cases, that was too hard or even impossible. Whenever a trade-off was needed, we always chose to remove \textit{false positives} entirely (i.e.~\textit{Dialyzer-style} certainty), even if that increased the number of \textit{false negatives}. In other words, \textbf{Hank} may not detect all the oxbow code instances, but every instance it reports \textit{is} an oxbow code instance.

\subsubsection{Preconditions}\label{preconditions}
Each rule in \textbf{Hank} assumes certain preconditions on the code that it is going to analyze. These assumptions limit the scope of application of the rule and present some scenarios in which the outcome of applying the rule is \textit{undetermined}. These limitations are listed in the module's documentation.
For instance, for the \textit{Unused Record Fields} rule, we have:

\begin{quote}
    This rule assumes that your code will never use the underlying tuple structure of your records directly.
\end{quote}

\subsubsection{Conditional Compilation}
It is also important to notice that \textbf{Hank} works directly on the code as it is written, i.e. without pre-processing. This is by design: on one hand, part of \textbf{Hank}'s job is to detect, for instance, unused macros (which is something that the Erlang pre-processor removes). On top of that, once again to avoid false positives, \textbf{Hank} will look through all the code that \textit{can} be compiled. In other words, if there is any way of compiling the code (say, defining a particular macro or assigning a particular value to it at compile time) that will include some pieces of code (e.g. a particular record definition), \textbf{Hank} will consider those ``used''. Check for instance, the following example:

\begin{lstlisting}
-define(SOME_MACRO, a:function(call)).

-ifdef(DEBUG).
another_function() -> ?SOME_MACRO.
-else.
another_function() -> ok.
-endif.
\end{lstlisting}

\verb`?SOME_MACRO` will be \textit{unused} unless the code is compiled with \verb`DEBUG` defined. But \textbf{Hank} will not report that macro as unused, because there is a way to compile the code in which it is actually needed and therefore it can not be safely removed.

\subsubsection{Unused Record Fields}
Originally, we did not want to consider record creation (i.e.~building a new instance of a record) as record usage, since assigning a value to a record field is not \textit{actually} using it, but the parser that \textbf{Hank} uses (based on the one provided by OTP) did not allow us to implement this rule that way because it does not differentiate record creation from record pattern-matching. In other words, all the expressions in the following function generate virtually indistinguishable syntax tree nodes in the part where they use the record:

\begin{lstlisting}
a_function(#a_record{a_field = AValue}) ->
  R = #a_record{a_field = AValue},
  #a_record{a_field = AValue} = R,
  #a_record{a_field = AValue}.
\end{lstlisting}

Lines 1 and 3 have patterns in them, while lines 2 and 4 create new records, but according to the parser, they are all nodes with \verb`type = record_expr`. So, to avoid false positives stemming from code like the one in lines 1 and 3, we had to consider all these nodes as record usage.

In addition, Erlang also allows another kind of special syntax for records to allow developers to work on multiple record fields at once:

\begin{lstlisting}
%% @doc Returns a record with all 'null's.
nullified() -> #a_record{_ = null}.
\end{lstlisting}

This is particularly useful when working with match expressions for \verb`ets` or \verb`mnesia`. In these scenarios, we just considered all the fields in the record as used.

\subsubsection{Unused Header Files}
Attributes \verb`include` and \texttt{in\-clude\_lib} both use relative paths that are affected by the list of include folders (provided with the \verb`-I` argument to the compiler) and the current version of each library in a release (folders in releases are called \verb`app_name-vsn`). When running \textbf{Hank}, we do not have any of that required information, so we cannot know for sure which file is each \verb`include` and \verb`include_lib` attribute referring to precisely.

Always intending to remove false positives, we decided to fail on the safer side and follow this algorithm when comparing the path of a header file against the path of one of the attributes:

\begin{enumerate}
    \item If both paths are the same, then they match (i.e.~the header file is used).
    \item Remove all relative path information from the include path (i.e.~remove \verb`../` and \verb`./`).
    \item If the largest of both paths ends with an exact match of the shorter one, then they match.
    \item Otherwise, they do not match.
\end{enumerate}

That algorithm would determine that \verb`a_header.hrl` and \verb`app/a_header.hrl` match, but \verb`app/other_header.hrl` and \verb`app/a_header.hrl` will not. Neither will \verb`app2/h2.hrl` and \verb`app/h2.hrl`.

The problem, of course, arises when a project has two apps, each one with a header file called \verb`h.hrl` and only one module has the following code:

\begin{lstlisting}
-include("h.hrl").
\end{lstlisting}

In this scenario, only one of the header files is actually used since no module includes the other one. But \textbf{Hank} cannot tell which of the header files is the one that is being included, and therefore it does not report the other as unused.

\subsubsection{Unused Configuration Parameters}
The main issue with this rule appeared when we tried to identify the files to analyze in search for configuration parameters. The obvious ones (i.e.~\verb`*.app.src`) were easy, but they did not cover much ground. In most projects, particularly those implementing large applications, some sort of \verb`sys.config` file is used. That is why we decided to consider all \verb`*.config` files as well.

This decision brought several side-effects with it. First, the structure of \verb`sys.config` files is not the same as the structure for \verb`*.app.src` files. Thus, the rule must be able to parse both, finding the \verb`env` configuration section in the application files and reading the entire file in the config ones.

Then, there is the problem that \verb`sys.config` files usually contain configuration parameters for applications that are not in the system (i.e.~dependencies). Luckily, \verb`rebar3` offers a way to get the list of OTP applications in the system. We used that list to filter the information included in config files.

Finally, not all config files are \verb`sys.config` ones. Some files may not even be parseable as Erlang terms. The rule just ignores them. Others are parseable but do not work as configuration for the run-time, most notably \verb`rebar.config` and \verb`elvis.config`. To deal with this, we included a list of known config files that the rule should ignore.

\subsubsection{Unnecessary Function Arguments}
There are two main scenarios where a positional argument for a function \textbf{must} be ignored and cannot be removed, and therefore it is not an instance of oxbow code: when the function implements a behavior callback, and when the function is a NIF stub.

\paragraph{Behavior Callbacks} When a module implements a behavior, it has to implement all non-optional callbacks defined by it, even if it has nothing to do for them. That usually leads to functions like this one:

\begin{lstlisting}
handle_cast(_, State) -> {noreply, State}.
\end{lstlisting}

The function has effectively nothing to do with the first parameter of that function, so it would seem unnecessary. But the behavior demands the implementation of \verb`handle_cast/2` not \verb`handle_cast/1`. Hence, the rule should not report this as an oxbow code instance.

Detecting functions that implement callbacks is not a trivial task. If the module defining the behavior is already compiled at the time of running the rule, \textbf{Hank} can collect the callback definitions by calling \texttt{module:behavior\_info(call\-backs)}. Nevertheless, if the module is not accessible (for instance, if it is part of the code that \textbf{Hank} has yet to analyze), \textbf{Hank} will need to find the callback definitions by itself in the module AST.

Moreover, not all callbacks for a particular behavior may be explicitly listed as \textit{callbacks}. An example of this is the \verb`gen_statem` behavior, which defines an arity for each of its state function callbacks, but not a name for those functions: the implementer can choose any name for those functions, as long as they have the proper arity and are exported.

To fail on the safer side, we added a list of \textit{known behaviors} to the rule (including, among others, \verb`gen_server` and \verb`supervisor`). For modules implementing those behaviors, \textbf{Hank} checks all functions except the expected callbacks. For all other modules implementing a behavior, this rule ignores all functions for the time being\footnote{In order to find a way to deal with this issue in a better way, we opened a ticket for the Erlang/OTP team~\cite{otp-ticket}.}

\paragraph{NIF Stubs}. When implementing a NIF (\textit{Native Implemented Function}), developers are required to create an associated Erlang module with the same interface but with all NIF functions being just stubs. This rule assumes that, following OTP recommendations, such stubs will just call \verb`erlang:nif_error/1,2` and ignore all their arguments. These arguments are not oxbow code instances, and to avoid reporting them as such, the rule ignores all functions that contain a call to \verb`erlang:nif_error/1` or \verb`erlang:nif_error/2`.

\subsection{Rule Exceptions}
Even after all the decisions we have described that have been taken to achieve the highest degree of certainty when detecting oxbow code instances, there will be instances in which the developers will like to preserve some particular oxbow code instances. It can be for documentation purposes, or because the particular project doesn't align with the preconditions of a rule, as described in section~\ref{preconditions}. To deal with these scenarios, \textbf{Hank} provides multiple ways to determine which warnings to ignore:

\begin{enumerate}
    \item Specifying particular instances to ignore or even entire files using the \verb`ignore` option in the \verb`hank` section of their \verb`rebar.config` files.
    \item Entire modules can be ignored by adding the attribute \verb`-hank ignore.` to them.
    \item Particular rules or particular oxbow instances can be ignored in a module using the \verb`hank` attribute with appropriate options, as seen below:
\begin{lstlisting}
-hank([{unused_record_fields,
          [ignored_record,
           {a_record, field_1},
           {a_record, field_2}
           ]}]).
\end{lstlisting}
This code will prevent \textbf{Hank}'s Unused Record Fields rule from emitting warnings about fields \verb`field_1` and \verb`field_2` in the record \verb`a_record`, or about any field in the record \verb`ignored_record`.
\end{enumerate}

\subsection{Performance}
As we showed in Fig.~\ref{fig:sequence}, \textbf{Hank}'s algorithm has two main steps:
\begin{enumerate}
    \item Parsing all the files, which is mainly impacted by the number of files to parse and their size.
    \item Running the rules on the ASTs, with its execution time derived from the number of rules and the complexity of their algorithms.
\end{enumerate}
Before releasing version \texttt{0.4.0} of \textbf{Hank}, we run some benchmarks for both of those stages. We compared the execution times of each stage, both running it in a parallelized way (using \verb`rpc:pmap/3`) and a sequential way (using list comprehensions). The projects we benchmarked were the same included on Table~\ref{tab:analysis}, and the results are presented in Table~\ref{tab:benchmark}\footnote{The benchmarks were executed in a Macbook Pro with a 2.6 GHz 6-Core Intel Core i7 processor and 16 GB of RAM and an SSD hard drive.}.

\begin{table*}
  \caption{Benchmarks for Hank Execution}
  \label{tab:benchmark}
  \begin{tabular}{lrrr}
  \toprule
                                 & \textbf{Erlang/OTP}   & \textbf{Kazoo}     & \textbf{MongooseIM} \\
  \midrule
  Size of project (in SLOC)      & $> 2M$       & $> 400K$  & $> 180K$  \\
  \midrule
  Sequential File Parsing        & $29042ms$    & $3457ms$  & $1735ms$  \\
  Parallel File Parsing          & $8751ms$     & $959ms$   & $490ms$   \\
  \midrule
  Sequential Rule Evaluation     & $150274ms$   & $6818ms$  & $2430ms$  \\
  Parallel Rule Evaluation       & $156136ms$   & $10678ms$ & $2788ms$  \\
  \midrule
  Total Sequential Runtime       & $181750ms$   & $10665ms$ & $4372ms$  \\
  Total Parallel Runtime         & $168057ms$   & $12101ms$ & $3635ms$  \\
  \midrule
  \textbf{Total Optimal Runtime} & $162424ms$   & $8360ms$  & $3393ms$  \\
  \bottomrule
  \end{tabular}
\end{table*}

We found that file parsing was greatly improved by using parallel execution, but rule evaluation was actually faster if performed sequentially. This may seem counter-intuitive, but it can be explained by the fact that there are very few rules, and none of them are bound by I/O or network processing. That is why the time spent by \verb`rpc:pmap/3` spawning processes, waiting for results, and collating responses outweighs the time spent actually running the algorithms.

With that consideration in mind, we came up with the \textbf{Optimal Runtime} shown as the last row of Table~\ref{tab:benchmark}, which is the one currently implemented in \textbf{Hank}: file parsing is run in parallel, but rule evaluation is performed sequentially.

\section{Related Work}

Code analysis and code quality improvement have long been dynamic research areas within software engineering research. Efforts have targeted from unused variables~\cite{Wand1999291} or dead-code elimination in recursive data~\cite{Liu2003221} to parallel programs~\cite{Knoop1998365}.

We find the most applied state of the art in this field to be language-specific, thus exploring the ways in which programming paradigm and language features manifest, specifically, as unreachable or dead code~\cite{Chen1998682}, precisely as we have done in our work. In many cases research is focused on the formulation, and sometimes formal verification of the detection/removal technique~\cite{Blech2005200}, but we find also empirical evaluations~\cite{Boomsma2012511} and tool implementations~\cite{Geneves2010305}.

As far as functional languages are concerned, most previous research has focused on typed languages~\cite{Damiani199983}, with the most recent efforts applying dependent data type for dead-code elimination~\cite{Xi1999228}. Techniques such as program transformation with dynamic rewrite rules attempt to overcome the context-free nature of rules~\cite{Bravenboer2006123}. Last but not least, partial dead code elimination~\cite{Knoop1994147}, i.e.~code that is only dead on some program paths, has also been proposed, but it requires second order effects to be removed, or else a limited scope (using program slicing~\cite{Bodik1997159}).

\subsection{Related Tools for Erlang}
\subsubsection{Other \texttt{rebar3} Plugins}
\textbf{Hank} is just one of the many plugins available for \verb`rebar3` that can increase maintainability. As stated in Sect.~\ref{unusedargs}, we actually \textit{assumed} that developers would run \textbf{Hank} together with some (if not all) other tools, and that is why we did not create rules to generate reports for aspects that are already covered by those publicly available tools. For reference, some of these tools and their relation with \textbf{Hank} are:
\begin{description}
    \item[Elvis] Elvis~\cite{elvis} is a linter: it reviews code style, like function naming, nesting level, line length, variable naming convention, etc. It works on a file-by-file basis, while \textbf{Hank} rules work on all the files of a project at once.
    \item[XRef] XRef~\cite{xref} is a cross-reference tool that can be used to find dependencies between functions, modules, applications, and releases. It is a great tool to detect dead code, but it cannot detect oxbow code instances since it works on compiled beam files.
    \item[Dialyzer] Dialyzer~\cite{dialyzer} is a static analysis tool that identifies software discrepancies using success typing. Just like XRef, it analyzes compiled (or at least fully-inter\-preted) code. Macros and records, among other classes of oxbow code, are virtually \textit{invisible} to Dialyzer.
\end{description}

\subsubsection{Other Erlang Tools}
Besides the ones that can be used via \verb`rebar3`, the Erlang Ecosystem provides several other useful tools that are in one way or another related to \textbf{Hank}. For instance, the Erlang Language Sever~\cite{erlangls} (\verb`erlang-ls`). ELS is a language server (based on the Language Server Protocol~\cite{lsp}) that provides language features (like auto-complete or auto-formatting) for the Erlang programming language. In principle, \verb`erlang-ls` could integrate with \textbf{Hank} to also provide automatic oxbow code detection, but the protocol works on a file-by-file basis so it cannot easily analyze whole systems at once. Nevertheless, the rules that can work per-file (e.g.~unused macros) are being ported to \verb`erlang-ls`, and the language server is already providing its own implementation of some of this functionality.

The Erlang Ecosystem has seen a number of static analysis and refactoring tools. From those still actively developed or maintained, \textbf{RefactorErl}~\cite{Bozo2011138,refactorerl} stands out. It is an open-source static source code analyzer and transformer for Erlang, developed by the Department of Programming Languages and Compilers at the Faculty of Informatics, Eötvös Loránd University, Budapest, Hungary.  So far, no analysis of possible collaboration between \textbf{RefactorErl} and \textbf{Hank} has been made, but this is certainly a great path for further investigation. An interesting idea would be to use \textbf{RefactorErl} to automatically \textit{fix} the warnings reported by \textbf{Hank}, or to use \textbf{RefactorErl} semantic analysis framework to extend the scope or overcome the current limitations of some rules.

\subsection{Related Tools for Other Languages}\label{otherlangs}
Within the Erlang Ecosystem, \textbf{Hank} is the only tool to detect oxbow code. There are no similar tools for other languages, like Elixir, LFE, or Gleam, and projects written in those languages cannot use \textbf{Hank}. A good path for further research would be to analyze such languages and the development of their corresponding oxbow code detectors. In doing so, our work can serve as inspiration on how to design them, and how to implement similar detection rules avoiding potential pitfalls.

Other communities do have similar tools. For instance, for code written in Python, there is \verb`vulture`~\cite{vulture}. Vulture finds unused code in Python programs. As opposed to \textbf{Hank}, Vulture developers decided to reduce the number of \textit{false negatives}, so it may emit warnings for code that is actually used. What they provided as a trade-off is a configurable level of confidence for each warning. It is a slightly arbitrary value that determines how confident Vulture is of a particular piece of code being dead, and the developers can filter Vulture output to just emit warnings with a minimum level.

Other tools, like \verb`phpdcd` for PHP~\cite{phpdcd} or \texttt{dead-code-de\-tect\-or}~\cite{dcdjava} for Java, have a somewhat different focus than \textbf{Hank}, and they mainly aim at finding unreachable or redundant code.

\section{Conclusions}
In this paper, we have addressed the problem of the identification of dead code, focusing mainly on \textbf{oxbow code}, in the context of the Erlang programming language. By analyzing several relevant open-source projects with many years of development, we showed evidence of the preeminence of this issue in large codebases.

As a solution to this problem, we have presented an open-source static analyzer, \textbf{Hank}, that can help to improve the maintainability of Erlang projects by allowing developers to detect and confidently remove unneeded portions of code. It is a \verb`rebar3` plugin that is easy to use, extensible, optimized for performance, and provides a high level of accuracy. We have described in detail how it is built, and discussed the reasons behind its design decisions. We believe \textbf{Hank} is ready for production usage, and that it should be included in the tool-chain of all well-maintained Erlang systems, together with Elvis, XRef, Dialyzer, and others.

Further work in \textbf{Hank} itself will be focused on the improvement of the existing rules, the creation of new ones, and the addition of new options that may facilitate its usage and integration even more, such as the optimization of file parsing and rule evaluation on unchanged files.

At the same time, the knowledge gained during the development of Hank could be put at service in the analysis of other languages in the Erlang Ecosystem, to determine how common it is for them to generate oxbow code instances, what kind of specific instances they may present, and how to best detect them. In cases where it is deemed suitable, the next step would be creating tools similar to \textbf{Hank} to detect and remove code that is no longer needed.

On a more abstract level, we intuitively know that reducing the general code surface of a project while not affecting its functionality improves maintainability. However, we do not have solid data on this matter. A future project may use \textbf{Hank} to verify how much faster or easier it is to work in a project once all instances of oxbow code have been removed.

\begin{acks}
We wish to acknowledge the support received from NextRoll and Fiqus, who devoted large chunks of their employees' paid work time to the development and maintenance of this open-source project and the writing of this paper, as part of their ongoing contributions to the Erlang Ecosystem Foundation.

We also wish to acknowledge the support received from the Centro de Investigación de Galicia (CITIC), funded by Xunta de Galicia and the European Union (European Regional Development Fund -- Galicia 2014-2020 Program), by grant ED431G 2019/01.
\end{acks}

\bibliographystyle{ACM-Reference-Format}
\bibliography{citations}
\nocite{easteregg}
\appendix

\end{document}